\documentclass[doublecol]{epl2} 
\usepackage{color}
\usepackage{graphicx}
\usepackage{amsmath, amsthm, amssymb}
\usepackage[colorlinks,citecolor=red,urlcolor=blue]{hyperref}
\newcommand{\be}{\begin{equation}}
\newcommand{\ee}{\end{equation}}
\newcommand{\bea}{\begin{eqnarray}}
\newcommand{\eea}{\end{eqnarray}}

\title{Phase transitions in systems of particles  with only hard-core interactions}
 
\author{Deepak Dhar\inst{1} \and R. Rajesh\inst{2,3} \and Aanjaneya Kumar\inst{1}} 
\shortauthor{D. Dhar, R. Rajesh and A. Kumar}
\institute{     
  \inst{1} Indian Institute of Science Education and Research, Dr. Homi Bhabha Road, Pashan,  Pune 411008, India\\            
  \inst{2} The Institute of Mathematical Sciences - C.I.T. Campus, Taramani, Chennai 600113, India\\
  \inst{3} Homi Bhabha National Institute, Training School Complex, Anushakti Nagar, Mumbai 400094, India}

\abstract{This article contains  our  comments and views on the status of the  current understanding  of phase transitions in systems with only hard-core interactions, based on our work in this area. The equation of state for the hard sphere gas in $d$-dimensions is discussed.  The  universal repulsive Lee-Yang singularities in the complex activity plane, and its relation to the directed and undirected polymer models are outlined. We also discuss orientationally disordered crystalline  mesophases, and some of their models.}

\begin{document}
\maketitle


\section{Introduction}

The change of state from the solid to the liquid to the gaseous state has been  known to humans since prehistoric times. Undoubtedly, the need to understand  this familiar, reproducible and yet  spectacular change  was the driving reason behind the interest in the study of phase transitions. The first scientific experimental investigations of phase transitions may be said to be  those by Cagniard de la Tours in 1822~\cite{cagniard}.  In the  last 200 years, phase transitions have been studied extensively, specially in statistical physics.  However, it is interesting that we still lack an exactly solved  model which shows solid to liquid to gas phase transitions.  While  Monte Carlo simulations have shown that assemblies of molecules with Lennard-Jones type pair-wise additive interactions, repulsive as short distances, and attractive at   larger distances  do show both  these transitions,  theoretically tractable models only show either one or the other transition, but not both!

\begin{figure}
\begin{center}
\includegraphics[width=0.8\columnwidth]{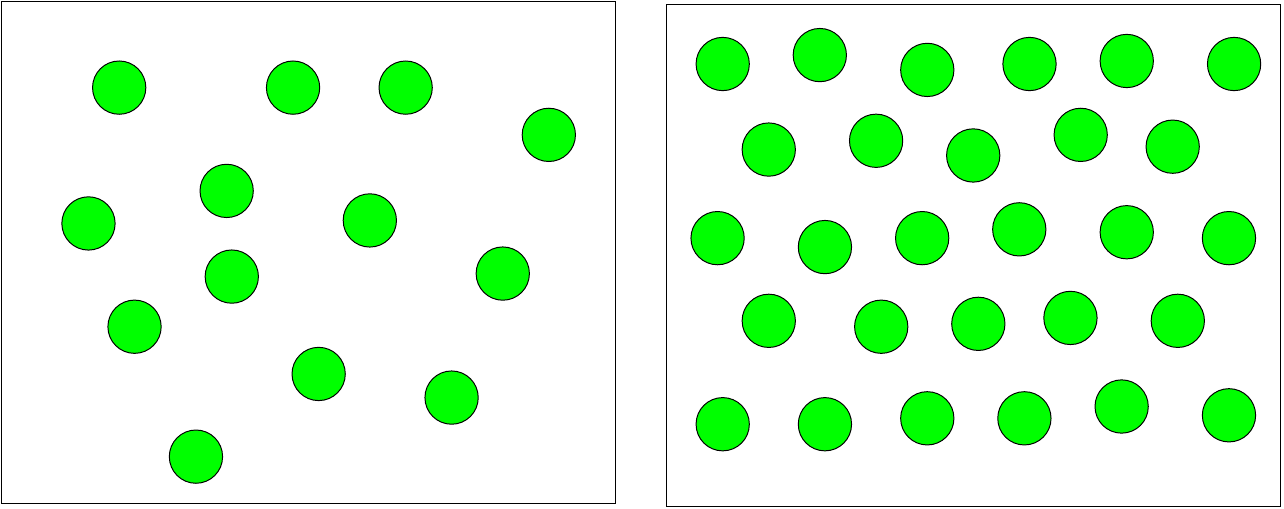}
\caption{ A schematic  two-dimensional representation of the low -density fluid and the high-density solid phases in a system   of hard spheres.}
\label{fig1}
\end{center}
\end{figure}

Since models with even the simple Lennard-Jones type pairwise interactions seem analytically intractable, the study of  a simpler model with the interaction simplified to a  purely repulsive power-law potential  $U(r) \sim (\sigma/r)^{n}$ seems useful,  despite the disadvantage  that  the liquid-to-gas transition cannot be obtained with purely repulsive potentials.   To ensure that the energy per molecule remains finite at non-zero densities, we restrict $n$ to be greater than $d$, the dimension of space.  In this case the equation of state becomes simpler, and the scaled pressure $ P$  becomes a function of a single scaling variable: $ P = f(\rho T^{-d/n}) $, where  $\rho$  is the number density, and $f(x)$ is a non-decreasing function of its argument $x$.  The limiting case of $n \to \infty$ is the hard sphere model, and the dependence on temperature completely drops out. This limit  is especially appealing, as here the phase transition is of purely geometrical origin: for hard spheres in a box, with   all allowed configurations equally likely,  there are densities $\rho_1$ and $\rho_2$, with $\rho_2 > \rho_1$, such that for densities below $\rho_1$, molecules have no long-range order, and the typical state of the system is `fluid-like', but for $\rho > \rho_2$,  translational invariance is spontaneously broken, and the centers of spheres form a hexagonal close-packed  lattice, as shown schematically in   Fig.~\ref{fig1}.

While  hard spheres cannot be said to be good models of molecules in real solids, it turns out that the attractive part of the interaction between them,  responsible for the binding energy in real solids, does not play a very significant role in the melting transition. The  binding energy per particle does change across the melting transition, contributing to  the latent heat of melting, but the changes in the  geometrical structure  of the simple solids (say, with one atom per cell) on melting  are fairly well captured in this simple model. Thus, the hard sphere model provides a simple qualitative explanation for the empirical observation that the entropy of fusion per atom,  for many simple materials,  is approximately $1$~\cite{wallace}.

\section{ Exactly soluble cases}

\begin{figure}
    \centering
    \includegraphics[width=0.6\columnwidth]{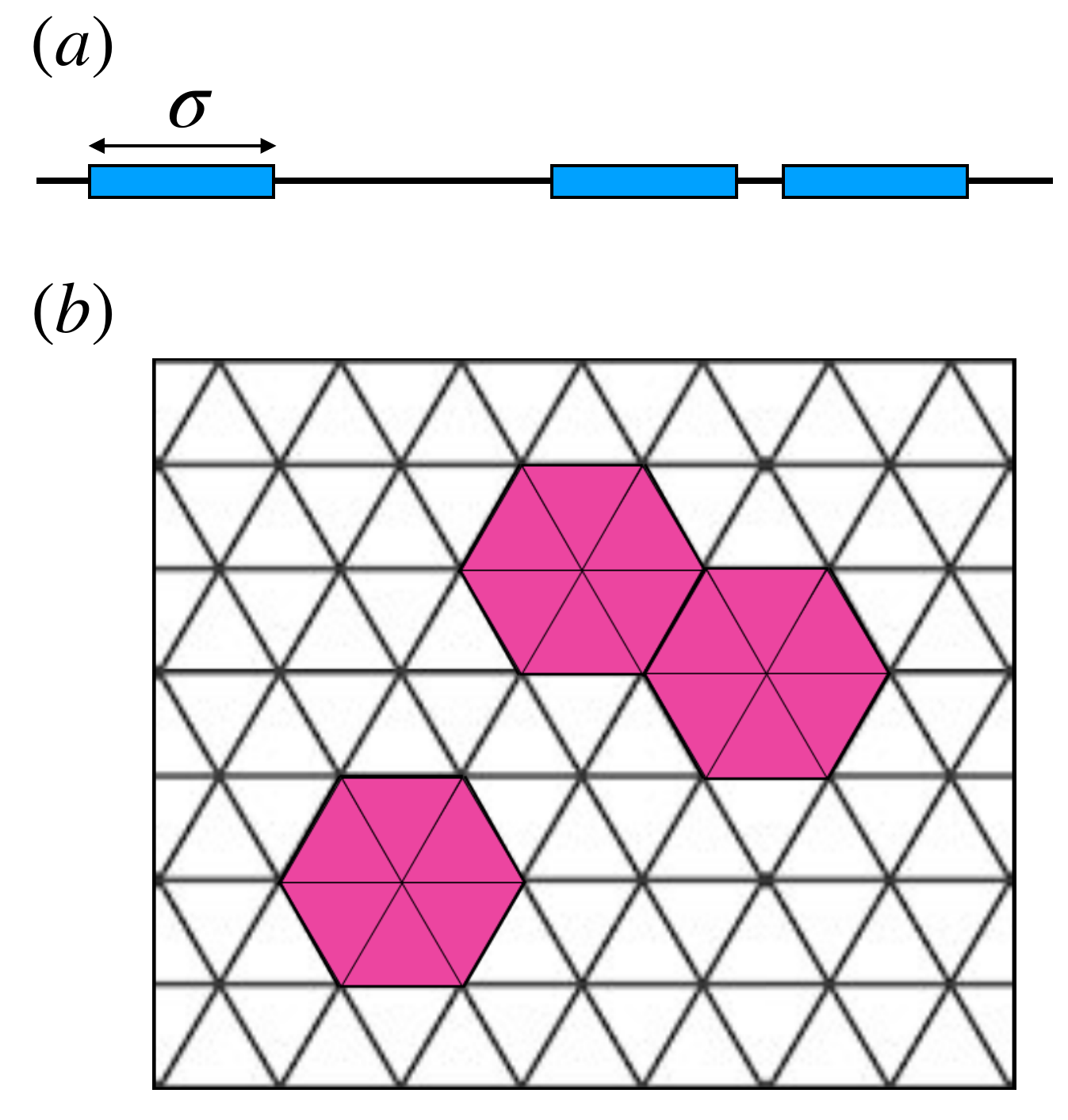}
    \caption{A schematic representation  of the two exactly solvable systems of hard-core particles. (a) Tonks gas, where hard rods of size $\sigma$ are placed on the 1D line. (b) Hard hexagons on the triangular lattice.}
    \label{fig:exact}
\end{figure}

Exactly soluble  models are very important for developing our understanding.  Unfortunately,  there are only  two known cases for hard-core particles: Tonks gas in one dimension, and Baxter's hard hexagon gas in two dimensions. For the Tonks gas  of hard core particles in one dimension [see Fig.~\ref{fig:exact}a], the equation of state in the continuum case is~\cite{tonks}
 \begin{equation}
 P =  \frac{k_B T \rho }{ 1 -\sigma \rho},
 \end{equation}
where $P$ is the pressure of the gas, $\rho$ is the number density per unit length, $T$ is the temperature, and $\sigma$ is diameter of the molecules. The problem of rods of length $k$ on a discrete lattice can also be solved \cite{shah2022phase}. 

The hard hexagon gas  on the triangular lattice [see Fig.~\ref{fig:exact}b], was solved by Baxter~\cite{baxter1,baxter2}. In the high density phase, one of the three sublattices of the triangular lattice is preferentially occupied, while in the low-density phase, all sublattices are equally occupied. The transition is continuous, and belongs to the universality class of the $3$-state Potts model, with the order parameter exponent $\beta = 1/9$, and correlation length exponent  $\nu =5/6$.  
Remarkably, in this case, there is a polynomial equation  that relates the density $\rho$ to the activity $z$,  of the form
\begin{eqnarray}
\nonumber
\rho ( 1 - \rho)^{11} -( 1-\rho)^5 P_1(\rho) z + \rho^2 ( 1- \rho)^2 P_2(\rho) z^2 \\
 -\rho^5 P_1(\rho) z^3 + \rho^{11} (1 -\rho) z^4 =0,
\end{eqnarray}
where $P_1(\rho) $ and $P_2(\rho)$ are explicit polynomials in $\rho$ with integer coefficients, of degree $7$ and $8$ respectively~\cite{joyce}.
 
\begin{figure}
\begin{center}
\includegraphics[width=0.8\columnwidth]{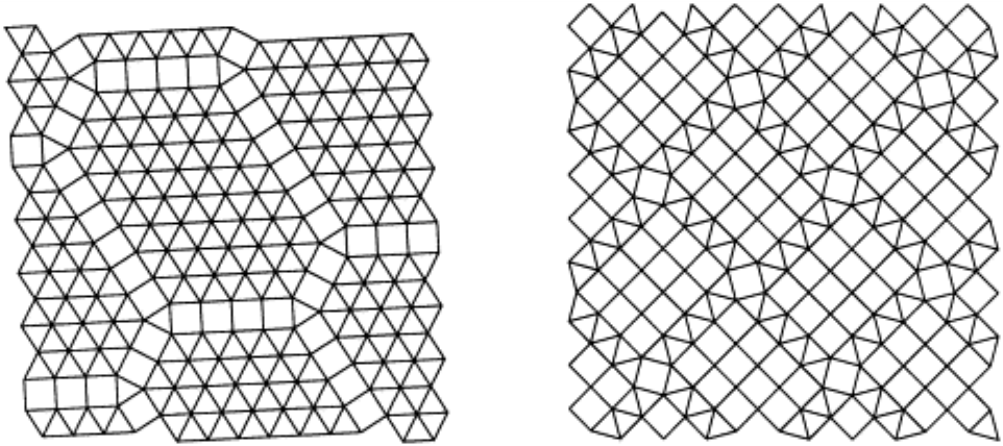}
\caption{ Examples of  covering of a plane by square and equilateral triangular tiles,  with majority of the tiles triangular(left)  and squares (right).  Figure taken from \cite{nienhuis}.}
\label{fig3}
\end{center}
\end{figure}

 A related problem that can be solved exactly is the case of complete tiling of a plane by square and triangular shaped tiles of equal edge length.  The configurations may be seen as defect lines of squares moving in a background of  triangles, or vice versa [see Fig.~\ref{fig3}]. The transfer matrix can be diagonalized using the  Bethe ansatz techniques,  to obtain  the entropy per unit area, as a function of fractional area covered by the triangles~\cite{nienhuis}. 

\section{ The hard-sphere equation of state and virial expansions}

\begin{figure}
\begin{center}
\includegraphics[width=0.75\columnwidth]{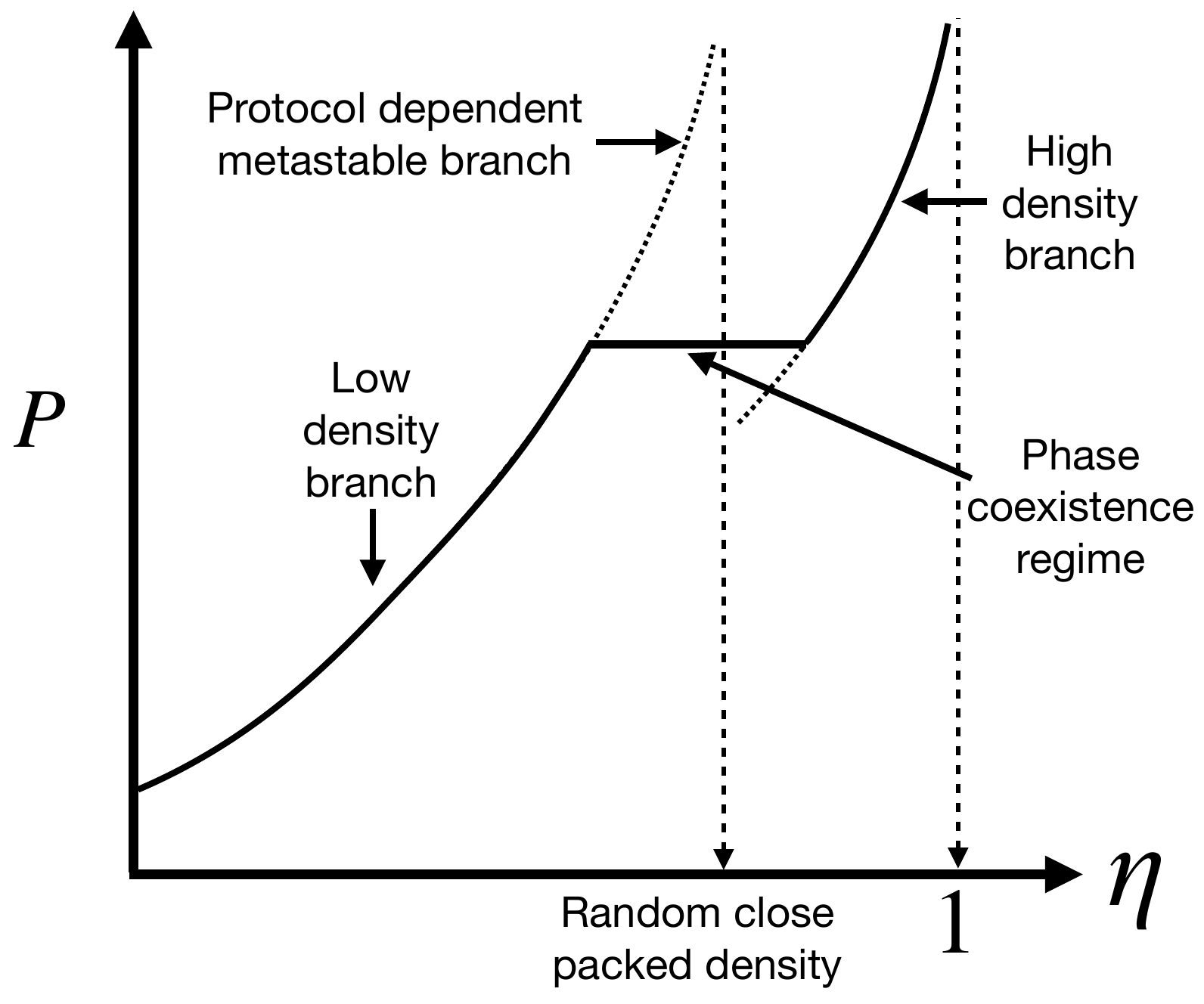}
\end{center}
\caption{Scaled pressure $P$ as a function of the scaled density $\eta$ of the hard-sphere system in three dimensions,  as seen in Monte Carlo simulations. The co-existence is for pressure $\approx 11.576$, for  the densities in the interval $[.492,.545]$.}
\label{fig:hardspheres}
\end{figure}

The expected qualitative behavior of the equation of state of hard spheres in $d=3$ is sketched in Fig.~\ref{fig:hardspheres}. There is a first order phase  transition from the liquid to the solid phase with a  jump in density. If  the liquid is compressed not too slowly experimentally, one finds that the pressure varies  continuously with density, and the low-density branch continues into a metastable glass-like branch (which is weakly dependent on the rate of compression). As the density tends to a value called the random closed packing density, denoted by $\rho_{rcp}$,  the pressure along this metastable branch is expected to tend to infinity. This branch  is shown by a dotted line in Fig.~\ref{fig:hardspheres}. Similarly, there is a metastable continuation of the high-density solid branch below the melting density. A good theory of hard spheres may be expected to give the equation of state not only along the true equilibrium curve, but also for the metastable branches!

One of the oldest  theoretical approaches to determining the equation of state of hard spheres is the virial expansion, which is a perturbation expansion for the pressure in powers of activity or density.  A large amount of work has been devoted to study of the equation of state of hard spheres in different dimensions. For example, Clisby and McCoy  calculated the first ten virial coefficients in dimensions $2$ to $8$~\cite{clisby}.  With increased computer power, and improvements in algorithms using Monte Carlo sampling, the computations have been pushed up to $64$ coefficients for dimension up  to $100$~\cite{zhang}. Using such series expansions, one can  try to find  an equation expressing the  pressure as a rational, or algebraic, function of density.  Mulero~\cite{mulero} discussed over a hundred different approximate equations of state that have been proposed over the years, for hard discs and hard spheres!

Unfortunately, this effort has had only very limited success.  First, the radius of convergence of the virial series is much larger than the density at which  the freezing transition is expected to set in, and the Mayer series seems not to detect the first order transition at all.  Second, most of the proposals, including the often used Carnahan-Starling equation of state~\cite{mulero}, do not have the expected analytical structure of the pressure as a function of activity $z$.   Some simplifications occur in the limit of large dimensions~\cite{wyler, kurchan}.  But the nature of dense phase (is it crystalline?) in the large-$d$ limit   is not  clear.

There is an interesting question related to the hard sphere equation of state: close to the close-packing limit, what is the dependence of the lattice constant of the periodic solid on the density $\rho$?  One may naively guess that the answer is  $\rho^{-1/3}$, in units of the  lattice constant for close packing. It turns out that the actual lattice constant is a bit less than this value, as it is entropically favored to divide the volume into $N'$ cells, with $N'$ bigger than the number of particles, and keeping a small concentration of these cells vacant. In fact, the optimum value of the fraction of cells kept vacant is of order $ \exp[ -const/\epsilon^2])$, where $\epsilon=1 - \rho$~\cite{stillinger}. The existence of such essential singularities rules out  the possibility that pressure  and density can be related by a simple polynomial  equation.

\section{ Connection to the Lee-Yang edge singularity, directed and undirected branched polymers}

In the Lee-Yang theory of phase transition in hard core systems (and other interactions may also be present),  the singularity of pressure as a function of the activity $z$ in the complex-$z$ plane, occurs because of the divergence of the potential at the zeroes of the grand partition function \cite{yang}. In simple cases, in particular for the ferromagnetic lattice discussed by Lee and Yang,  these zeroes condense into lines of zeroes in the thermodynamic limit, but  even for  the `next-simplest'  case of the  nearest neighbor exclusion gas on the square lattice, there is numerical evidence of  an areal density of zeroes in a region of the complex plane~\cite{assis}.   

The density of zeroes at the end points of  lines of zeroes   has a power-law form that is universal, and independent of the details of the model.  The density of zeroes near an end-point  of a line of zeroes $z_c$ varies as $|z -z_c|^{\sigma(d)}$, and    is referred to as  the universal Yang-Lee edge singularity~\cite{lai, cardy}.  The value of $\sigma$ depends only on the dimension of space $d$, and  is known to be exactly  $-1,-1/2, -1/6$  for $d=0, 1, 2$ respectively. For $d \geq 6$ it takes the mean field value  $1/2$ \cite{lai}.   Correlation functions of the system at the complex value of activity $z = z_c$ show critical power law behavior. Unlike the usual critical phenomena, here all critical exponents can be expressed in terms of the single critical exponent $\sigma(d)$.

\begin{figure}
\begin{center}
\includegraphics[width=0.75\columnwidth]{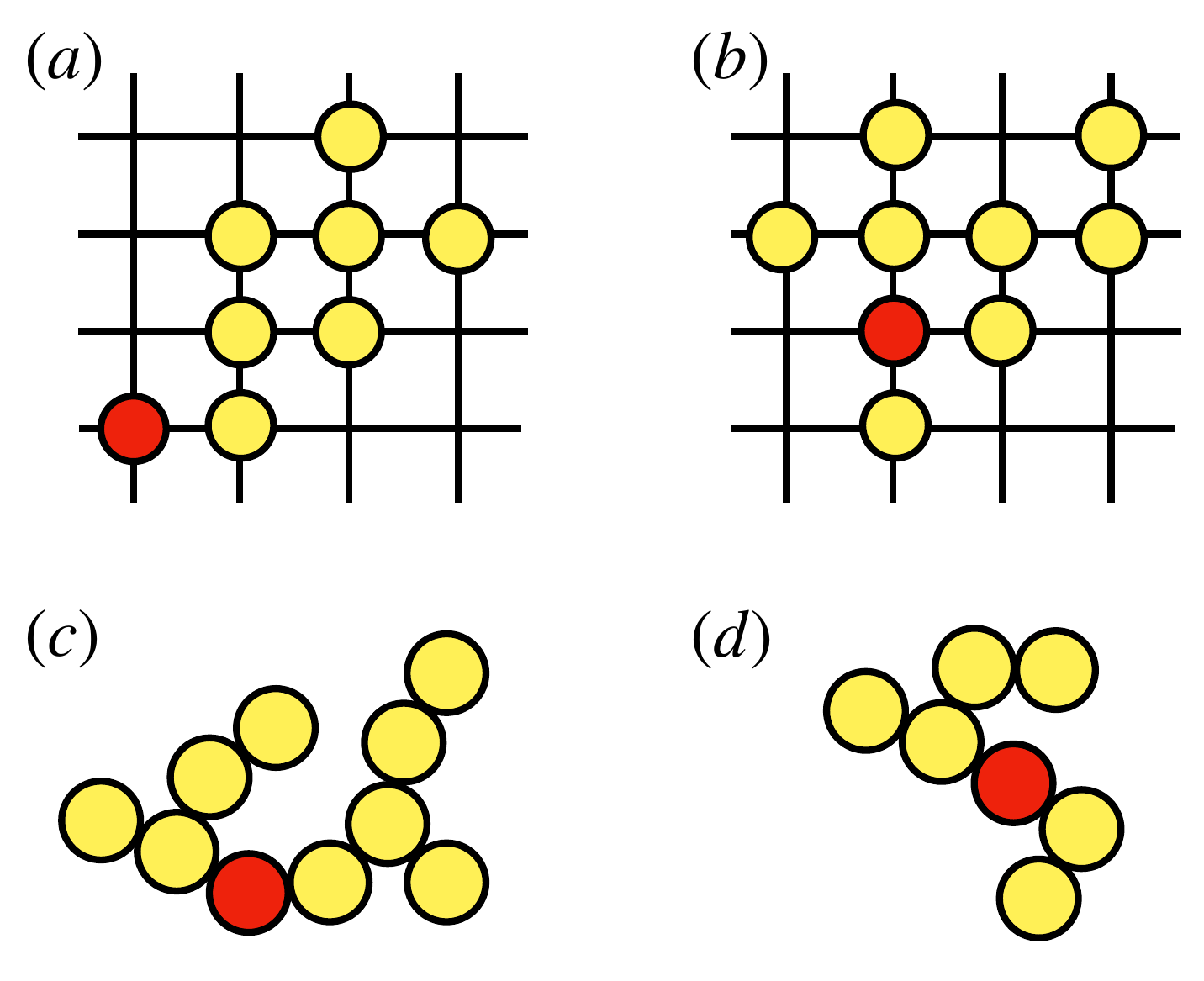}
\caption{ (a) A directed animal on the square lattice (b) An undirected animal on the square lattice (c) A directed cluster of touching discs with no loops on the $2D$ plane  (d) An undirected cluster of discs with no loops.}
\label{fig:animals}
\end{center}
\end{figure}

The study of the critical point  at the Yang-Lee  edge singularity  is complicated by the fact that because of negative weights of configurations, properties like convexity of free energy, and equivalence of ensembles do not hold.  Fortunately, the problem of Yang-Lee edge singularity is equivalent to the problem of the enumeration of branched polymers, directed, and undirected, in which all weights are positive. The definition of directed  and undirected polymers, on the lattice, and on the continuum are indicated in Fig.~\ref{fig:animals}. 
Let $A_{d,bch}(n)$ be the number of distinct directed animals on the $d$-dimensional body-centered hypercubical lattice (bch), with origin as the root, and $A_{d,bch}(x)$ is the corresponding generating function.  It was shown in Ref.~\cite{dhar} that 
\begin{equation}
A_{d,bch}(x) = -\rho_{(d-1),bch} \left( z = -\frac{x}{1-x} \right),
\end{equation}
where $\rho_{(d-1),bch}(z) $ is the density of a nearest neighbor exclusion lattice gas with activity $z$ on a $(d-1)$-dimensional body centered hypercubical lattice. 

There is a similar result for the directed branched polymer (DBP) clusters on the $2$-d plane (here the cluster is a set of touching discs, with no loops allowed). The weight of a graph involves sum over the different angles allowed, with the no loops constraint.  One finds  the exact expression for the sum of all sum weights of directed branched polymers~\cite{imbrie2003}
\begin{equation}
\rho_{d}(z) = -Z_{d+1,DBP}(-z),
\end{equation}
where $\rho_d(z)$ is the density of the hard sphere gas in $d$ dimensions with activity $z$, and $Z_{d+1,DBP}(-z)$ is the partition function of directed branched polymers in $(d+1)$ dimensions. 

Even more remarkably, the partition function for undirected branched polymers in $(d+2)$ dimensions is related to the $\rho_d(z)$ by the simple equation \cite{imbrie2003}
\begin{equation}
\rho(z) = - 2 \pi Z_{d+2, BP} \left( - \frac{z}{2 \pi} \right).
\end{equation}

\begin{figure}
\begin{center}
\includegraphics[width=0.8\columnwidth]{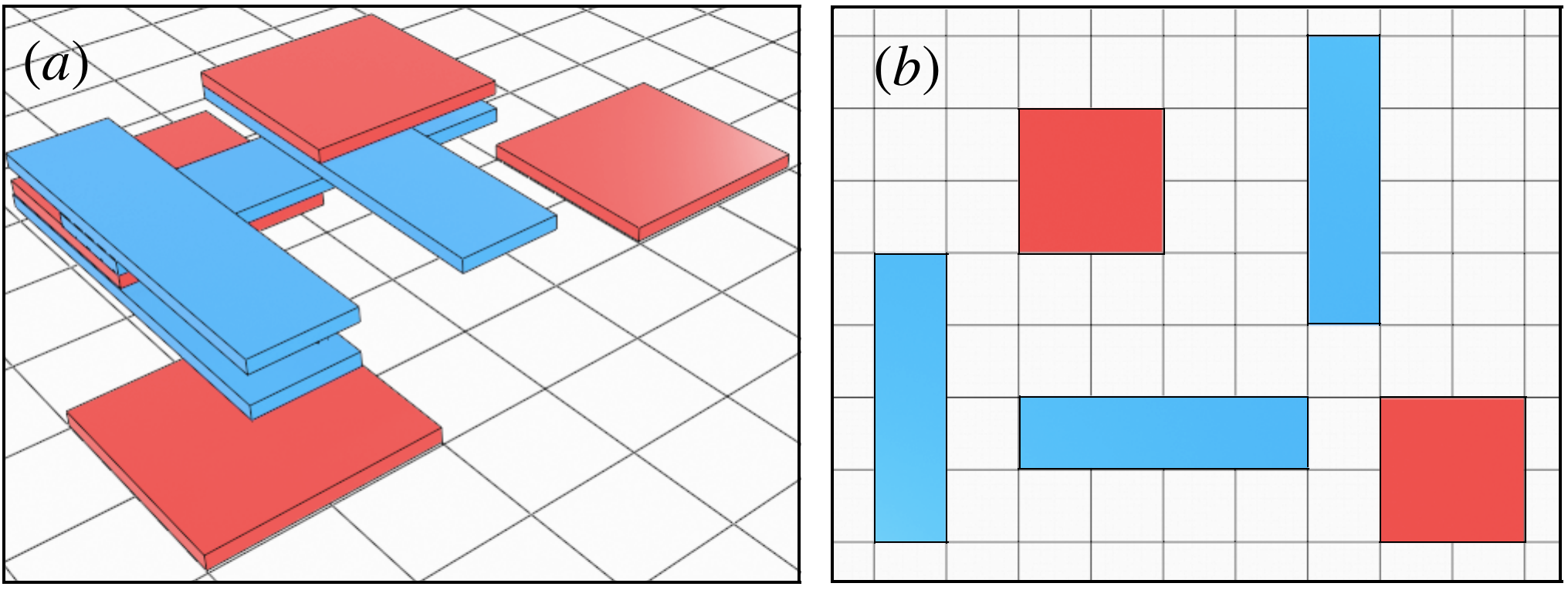}
\caption{ (a) A heap made of two different types of pieces. Pieces are on different heights. Each piece falls down till it can not fall any further under gravity, and is supported by a piece below. The figure depicts one heap which has a single piece at the lowest level, and another independent piece lying disconnected from the heap. (b) A collection of non-overlapping pieces on the 2D square lattice.}
\label{fig:heaps}
\end{center}
\end{figure}

There is a remarkable extension of the result for directed branched polymers to enumeration of heaps, made of any number of pieces of different shapes, each with a different activity $z_i$ for the $i$-th species~\cite{viennot}. One can then form the grand partition function $\mathcal{Z}(\{z_i\})$ in a finite volume $V$,  where these particles are non-overlapping. Then the density of the $i$-th type of particle in this ensemble is obtained by taking a derivative of this partition function. This can be shown to be exactly equal to the generating function of  heaps  made of all kinds of pieces and of arbitrary height, that are supported at the bottom by a single piece of the $i$-th species, for all $i$ ( see Fig.~\ref{fig:heaps}). 

We note  that only a few other examples of  dimensional reduction are known outside the context of hard core models (or their simple extensions), i.e. disorder solutions~\cite{rujan}, and models with supersymmetry \cite{parisi1979random}. A better understanding of its underlying mechanism could help  in the solution of other problems. 

\section{An alternative approach to the equation of state of systems of hard particles}

As noted above, for a given shape of hard particles,  finding the exact equation of state has been possible only in few cases. It then seems desirable to try to find an  approximate equation of state. Typically, such an equation should have a functional form that is  consistent with known results, and contain some adjustable parameters, whose values can be fixed by fitting to  experimental data  about the specific system.

In particular, we expect that the equation would satisfy the following conditions: (a) The pressure is an increasing function of density, with  expected behavior in the low- and high-density limits. (b) It shows the observed phases and phase transitions as the density is increased. (c) The
pressure as a function of density shows Yang-Lee edge singularity with the correct exponent in the complex activity plane. (d) The equation is consistent with the known values of coefficients in the low- and high- density expansions. 

We note that  all the equations of state proposed so far fail to satisfy at least one of these good behavior conditions. Given the problems with the virial expansion approach to constructing the equation of state, it seems worthwhile to try a different approach. We propose here  that we start with some parameterization  of the density of zeroes of the grand partition function in the complex plane, and use it to build the equation of state.

\begin{figure}
\begin{center}
\includegraphics[width=1.65in]{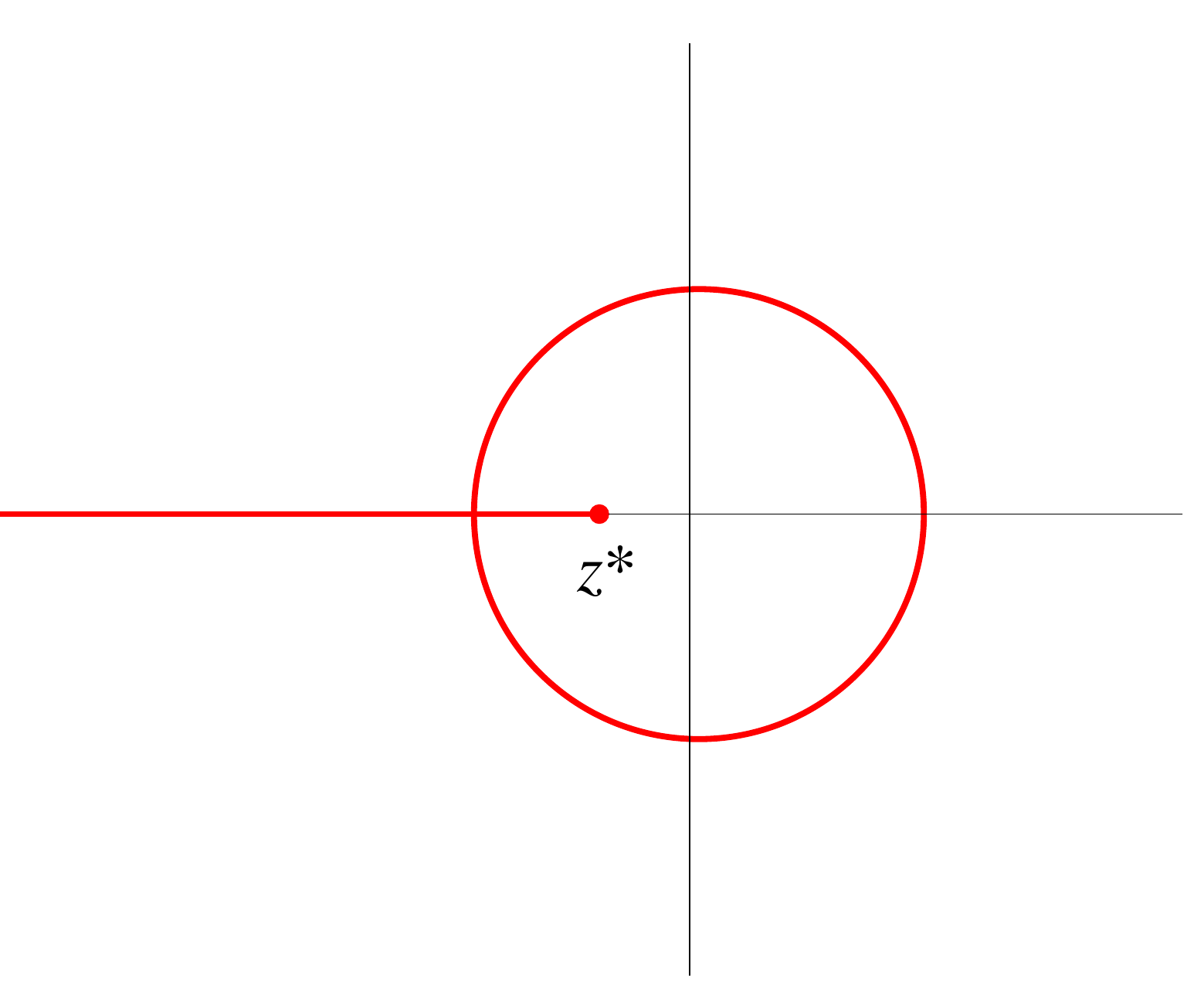}
\caption{A schematic  diagram showing the essential elements of the  zeroes in the complex activity plane for the hard sphere equation of state: a line on the negative real axis up to point $z^*$, which is the closest singularity of the partition function to the origin, and a line of zeroes forming a closed loop, taken to a circle for simplicity.}
\label{fig:ly-zeroes}
\end{center}
\end{figure}

In the simplest setting, the density of zeroes has two main components: A line of zeroes on the negative real axis, up to the point $z = z^* <0$ with density $f_1(z)$, and a closed loop of zeroes, which we take to be circle of radius $b$, with density of charges given by $f_2(\theta)$ along the polar direction $\theta$   [see Fig.~\ref{fig:ly-zeroes}].   For example, one can choose 
\bea
f_1(z) &=& \exp( - a z) |z - z^*|^{\sigma} \sum_{r=0}^{\infty} C_r z^r, \\
f_2(\theta) &=& B,
\eea
where $z^*$ is the nearest singularity to the origin on the negative real axis. 
Then the pressure $P(z)$ as  the electrostatic potential produced by this density of charges a piece-wise  analytic function of $z$.  Here, the loop of zeroes gives the observed first order transition.  The constants $b$ and $B$ determine the position and strength of the discontinuity in density.  In cases with more than one phase transition, one will need more that one closed loop. 

For this choice of $f_2$, as the electric field inside a uniformly charged ring is zero, the virial coefficients in the series for density as a function of $z$,  for $|z|<b$, are completely independent of $b$ and $B$.  Adjusting the coefficients, $C_r$, we can match as many virial coefficients as we wish. 
 
Now, consider the metastable branch. For our choice of $f_2$,  the analytic continuation of the low-$z$ branch  is obtained by  the same function by setting  $B=0$. Then, we find  that for a given activity $z$, the density in the  analytically
-continued metastable   is less than  the value in the stable high-density   for same $z$ by a constant amount $B$.  If the  change in density across the melting transition is denoted by $\Delta \rho (= B)$, we find 
\begin{equation}
\rho_{rcp} = \rho_{max}  -\Delta \rho.
\end{equation} 
This is in a qualitative agreement with the observed values for hard spheres in three dimensions.   A  better agreement can be achieved with   a more calibrated choice of $f_2$. The important point is that we are able  to address  questions regarding the metastable branch also within this scheme. 

For a given density of zeros, the electrostatic problem of finding the electric fields at different points in space for a given density of charges is straightforward. If the density of charges has an exponential tail in the negative real line, it is easily seen that the electric field at large positive $z$ will be a sum of the usual multipole expansions of the form $|z|^{-n}$, but also have corrections of the form $\exp(-|z|)$ corresponding to the fractional charge that is outside the disc of radius $|z|$. Thus, in this approach, we are able to incorporate the essential singularity expected   in the equation of state at large $|z|$ discussed above.

\section{Orientationally disordered crystalline phases}

It is important to note  that  the transition from a  low temperature crystalline phase  showing  broken translational and rotational invariance, to the higher temperature isotropic liquid phase is not always a single phase transition,  and often involves  one or more intervening  mesophases. These mesophases may be divided into two  broad types:  plastic solids and liquid crystals. In the former, there is a long-range   three-dimensional lattice  crystalline order, but within the unit cell, there is only partial order in the orientations of molecules.  In the latter, a three-dimensional crystalline lattice is  absent, but the system shows some orientational order.  
The former has been variously called  the rotator phase,  or the plastic crystal phase. Our preferred nomenclature  is  \emph{orientationally disordered crystals}  (ODCs). While the liquid crystalline phases are well studied in physics literature~\cite{degennes,1992-chandrasekhar}, ODC phases have not  received as much attention from  the statistical physics community. One of the aims of this article is to  highlight this interesting problem.

The ODC phases are seen in very simple  molecular solids, for example,  hydrogen chloride (HCl) and methane (CH$_4$).    In fact, the first mention of the idea of the plastic crystal phase was proposed by Simon and von Simson in the context of HCl a century ago in 1924 \cite{hcl}.  For example, when liquid methane is cooled,  it  first freezes into a state where the carbon atoms are arranged in a fcc lattice, but the molecules are  free to rotate without much steric hindrance from neighboring atoms, and orientationally disordered. Currently ODCs of somewhat complicated organic and ionic compounds are being investigated intensively because  of their technological interest  in  applications ranging from  batteries~\cite{batteries},  refrigeration~\cite{refrigeration}, drug-delivery~\cite{drug-delivery}, opto-electronic~\cite{opto} and piezo-electrical devices~\cite{piezo}.  For a recent review of applications of ODCs, see~\cite{Das}. 

\begin{figure}
    \centering
    \includegraphics[width=0.8\columnwidth]{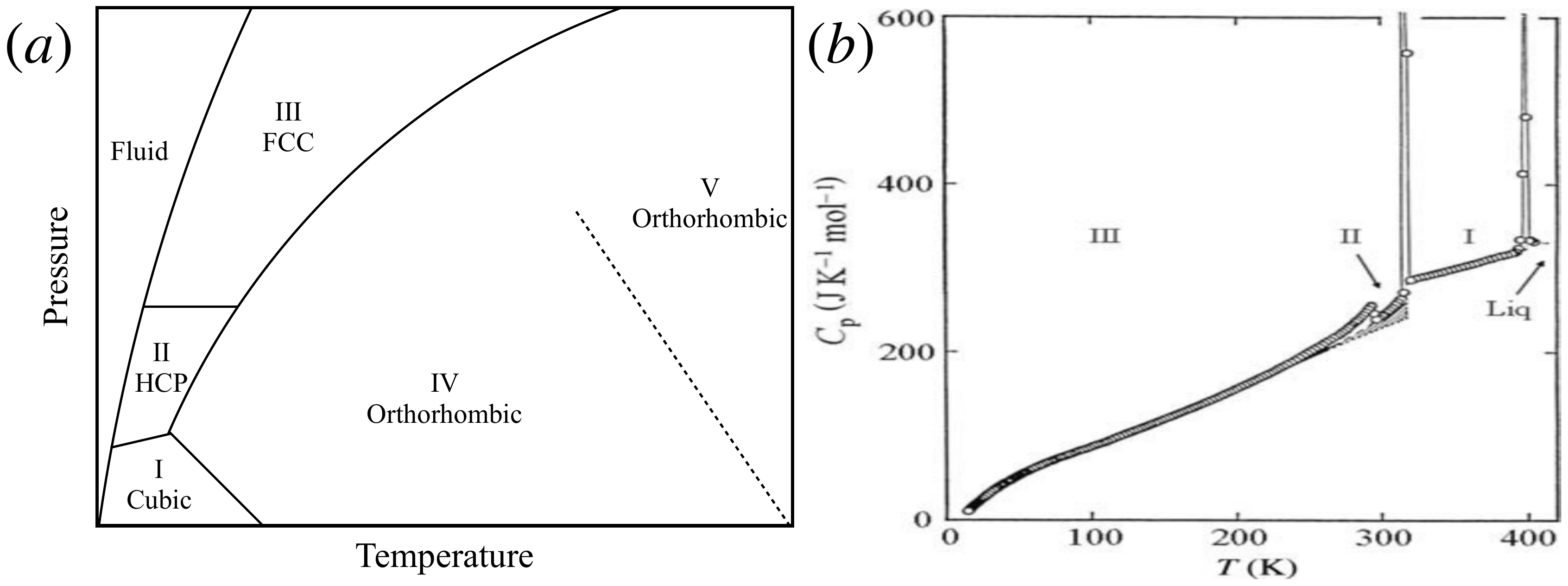}
    \caption{(a) A schematic phase-diagram of ammonia depicting the many  solid phases. The phases II and II are ODCs  [adapted from Ref.~\cite{ammonia}]. (b) Molar heat capacities of formylferrocene as a function of temperature, taken from Ref.~\cite{kaneko}. }
    \label{fig:odc}
\end{figure}

As an example of the rich phase structure of ODCs, we show the phase diagram of ammonia~\cite{ammonia} in Fig.~\ref{fig:odc}a. We emphasize that at present, there is no systematic way to  predict  which phases would be realized in a given material, and when there would be a sharp  phase  transition, as opposed to a crossover behavior under a change of the control parameters.  

The phase transitions from one ODC phase to another  may be first order or continuous.  As an example, in Fig.~\ref{fig:odc}b, we show the molar heat capacity as a function of temperature in formylferrocine. In this case, three phases are distinguishable, labeled here I, II and III.  Clearly, the transition from phase I to II seems to be first order, whereas the transition from II to III appears to involve only a finite jump in the specific heat.

Lattice models of hard core particles are more tractable theoretically. They are also more efficiently  simulated with cluster algorithms~\cite{kundu2013nematic,jaleel2021rejection} making it possible to sample even the fully packed limit.  There are many shapes for which the phases and phase diagrams have been obtained: tetrominoes, squares, hexagons, long rods, rectangles, triangles, Y shaped molecules etc (see \cite{barnes,2018-pre-mnr-phase,shah2022phase,rajesh2} and references cited therein). As simple examples of lattice models that show ODC phases, we note that the high-density phase of $2 \times 2k$ rectangles, with $k >1$ on a square lattice~\cite{rectangles,kundu}, and $ 2 \times 2 \times 1$ plates on a cubic lattice~\cite{geet} show crystalline (sub-lattice) order, but no orientational order.   
 
For hard core models, an open question is predicting the sequence of different phases, given that we know the phases, with increasing density. For example, for $2\times 2 \times 2$ cubes,  translational symmetry is broken in $0, 1, 3, 2$ directions in increasing order of density, showing no obvious pattern~\cite{rajesh1}.  

\begin{figure}
\begin{center}
\includegraphics[width=1.65in]{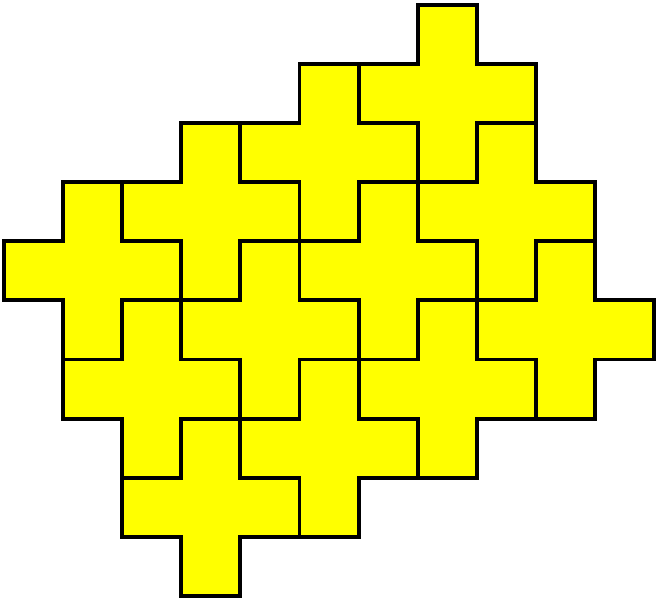}
\caption{ A tiling of the square lattice using cross-shaped pentamers}
\label{fig:cross}
\end{center}
\end{figure}

For a given shape, the first question is the value of maximum packing density. If the maximally packed state has zero entropy per site, we may have a breaking of translational and/or rotational   lattice symmetries.  Then, one may be able to show,  using a Peierls-type argument that this lattice structure is stable against introduction of a small concentration of vacancies.  For example, the cross shaped pentamer can achieve full packing, and there are 10 different fully packed states possible, for different boundary conditions [Fig.~\ref{fig:cross}]. 

Even when  the maximally packed state has zero entropy per site, if there is  a sliding instability \cite{runnels1965hard}, one could get a   partial restoration of translational symmetry, and the state of the system could show  columnar order, or layering~\cite{rajesh1, nath2016high,mazel2023hard}. Also, even if the maximally packed state has non-zero entropy per site, there may still be sublattice ordering, or other forms of order (e.g. hard cubes). In this case, the high density expansion involves fractional powers of  $z^{-1}$. 
Sometimes, one of the many ground states at full packing is selected on introducing vacancies  by the order-by-disorder mechanism. For example, in the case $ 2 \times 2 \times 2$ cubes on a cubic lattice~\cite{rajesh1}.

\begin{figure}
\begin{center}
\includegraphics[width= 2 in]{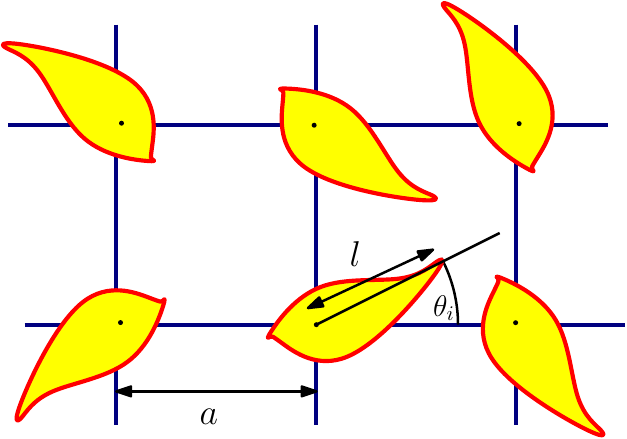}
\caption{ A configuration of a model of hard rotors pivoted to lattice sites. Taken from \cite{odc2}. }
\label{fig:rotors}
\end{center}
\end{figure}

\begin{figure}
\begin{center}
\includegraphics[width = 1.65 in] {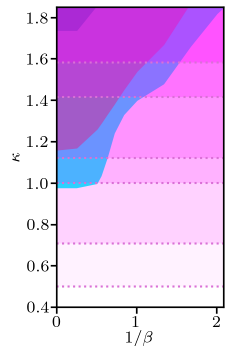}
\caption{Many phases seen in the rotors model, when the rotors are thin rods pivoted at mid points on  a square lattice. Here $\kappa$ is the ratio of the rod lengths to the lattice spacing, and $\beta$ is the inverse temperature, with all overlaps having  finite energy cost $1$. Different phases are shown in different colors.  Taken from Ref.~\cite{odc1}.}
\label{fig:many}
\end{center}
\end{figure}

Another class of models is where we consider a fully occupied lattice with a non-spherical molecule at each lattice position, but with orientational  degrees of freedom, and hard core interactions [Fig. \ref{fig:rotors}]~\cite{runnels1, runnels2, odc1,odc2}. These simple models can show a very large number of phases and phase transitions [Fig.~\ref{fig:many}].  

For these models, exact results are possible for some range of densities. It was shown in Ref.~\cite{odc2}
that  in some range of densities when the at-most one overlap condition holds the Mayer cluster expansion simplifies enormously:  the model is equivalent to monomer-dimer models, and   one can determine the orientation distribution function~\cite{odc2}. Also, even outside of  this range, one can determine the singularity structure of the orientation distribution function~\cite{odc3}.

\section{Summary and concluding remarks}

The existing physics literature on phase transitions is mainly devoted to understanding the critical phase transitions, and the critical exponents. Undoubtedly, the renormalization group approach to understand  critical  phase transitions has been a major achievement of the twentieth century. However, other equally basic qualitative questions about phase transitions such as, for a given material, what are the phases possible, and their relative arrangement in the phase diagram have not received as much attention. In particular, the multitude of ODC phases seen the phase diagram of even simple materials still lack explanation. 

For the hard sphere problem, progress in  obtaining the equation of state using the standard virial expansion technique seems to be difficult, and we have suggested an alternate approach starting with the density of zeroes in the complex plane. We also emphasize that dimensional reduction is a very powerful technique for obtaining exact solutions, and is a  promising direction for further study.


\acknowledgments{The authors thank M. Barma and K. Ramola for their critical comments on the manuscript. DD's work was partially supported by a Senior Scientist Fellowship by the National Academy of Sciences of India. AK acknowledges the Prime Minister's Research Fellowship of the Government of India for financial support. }

\bibliographystyle{eplbib}

\begin{thebibliography}{10}
\expandafter\ifx\csname url\endcsname\relax\def\url#1{\texttt{#1}}\fi

\bibitem{cagniard}
\Name{Berche B., Henkel M. \and Kenna R.} \REVIEW{J. Physical
  Studies}{13}{2009}{3201 [arXiv: 0905.1886]}.

\bibitem{wallace}
\Name{Wallace D.~C.} \REVIEW{Proceedings of the Royal Society of London. Series
  A: Mathematical and Physical Sciences}{433}{1991}{631}.

\bibitem{tonks}
\Book{$1$-dimensional hard rods}.
\newline\url{http://www.sklogwiki.org/SklogWiki/index.php/1-dimensional_hard_rods}

\bibitem{shah2022phase}
\Name{Shah A., Dhar D. \and Rajesh R.} \REVIEW{Physical Review
  E}{105}{2022}{034103}.

\bibitem{baxter1}
\Name{Baxter R.~J.} \REVIEW{J. Phys. A}{13}{1980}{L61}.

\bibitem{baxter2}
\Name{Baxter R.~J.} \Book{Exactly solved models in statistical mechanics}
  (Academic Press, London) 1982.

\bibitem{joyce}
\Name{Joyce G.} \REVIEW{J. Phys. A}{21}{1988}{L983}.

\bibitem{nienhuis}
\Name{Nienhuis B.} \REVIEW{Phys. rep.}{301}{1998}{271}.

\bibitem{clisby}
\Name{Clisby N. \and McCoy B.~M.} \REVIEW{J. Stat. Phys.}{122}{2006}{15}.

\bibitem{zhang}
\Name{Zhang C. \and Pettitt B.~M.} \REVIEW{Molecular Physics}{112}{2014}{1427}.

\bibitem{mulero}
\Name{Mulero A., Galan C. \and Cuadros F.} \REVIEW{Physical Chemistry Chemical
  Physics}{3}{2001}{4991}.

\bibitem{wyler}
\Name{Wyler D., Rivier N. \and Frisch H.} \REVIEW{Physical Review
  A}{36}{1987}{2422}.

\bibitem{kurchan}
\Name{Kurchan J., Parisi G. \and Zamponi F.} \REVIEW{Journal of Statistical
  Mechanics: Theory and Experiment}{2012}{2012}{P10012}.

\bibitem{stillinger}
\Name{Stillinger~Jr F.~H., Salsburg Z.~W. \and Kornegay R.~L.} \REVIEW{J. Chem.
  Phys.}{43}{1965}{932}.

\bibitem{yang}
\Name{Lee T.-D. \and Yang C.-N.} \REVIEW{Physical Review}{87}{1952}{410}.

\bibitem{assis}
\Name{Assis M., Jacobsen J., Jensen I., Maillard J. \and McCoy B.}
  \REVIEW{Journal of Physics A: Mathematical and
  Theoretical}{47}{2014}{445001}.

\bibitem{lai}
\Name{Lai S.-N. \and Fisher M.~E.} \REVIEW{The Journal of Chemical
  Physics}{103}{1995}{8144}.

\bibitem{cardy}
\Name{Cardy J.} \REVIEW{arXiv preprint arXiv:2305.13288}{}{2023}{}.

\bibitem{dhar}
\Name{Dhar D.} \REVIEW{Physical Review Letters}{51}{1983}{853}.

\bibitem{imbrie2003}
\Name{Brydges D.~C. \and Imbrie J.~Z.} \REVIEW{Annals of
  mathematics}{}{2003}{1019}.

\bibitem{viennot}
\Name{Viennot G.~X.} \Book{Heaps of pieces, i: Basic definitions and
  combinatorial lemmas} in \Book{Combinatoire {\'e}num{\'e}rative} (Springer)
  1986 pp. 321--350.

\bibitem{rujan}
\Name{Rujan P.} \REVIEW{Journal of statistical physics}{49}{1987}{139}.

\bibitem{parisi1979random}
\Name{Parisi G. \and Sourlas N.} \REVIEW{Physical Review
  Letters}{43}{1979}{744}.

\bibitem{degennes}
\Name{de~Gennes P.~G. \and Prost J.} \Book{The physics of liquid crystals}
  Vol.~83 (Oxford university press) 1995.

\bibitem{1992-chandrasekhar}
\Name{Chandrasekhar S.} \Book{Liquid crystals} (Cambridge University Press,
  Cambridge) 1992.

\bibitem{hcl}
\Name{Simon F. \and Simson C.~v.} \REVIEW{Zeitschrift f{\"u}r
  Physik}{21}{1924}{168}.

\bibitem{batteries}
\Name{Pringle J.~M., Howlett P.~C., MacFarlane D.~R. \and Forsyth M.}
  \REVIEW{Journal of Materials Chemistry}{20}{2010}{2056}.

\bibitem{refrigeration}
\Name{Li B., Kawakita Y., Ohira-Kawamura S., Sugahara T., Wang H., Wang J.,
  Chen Y., Kawaguchi S.~I., Kawaguchi S., Ohara K. \etal}
  \REVIEW{Nature}{567}{2019}{506}.

\bibitem{drug-delivery}
\Name{Shalaev E., Wu K., Shamblin S., Krzyzaniak J.~F. \and Descamps M.}
  \REVIEW{Advanced Drug Delivery Reviews}{100}{2016}{194}.

\bibitem{opto}
\Name{Sun Z., Chen T., Liu X., Hong M. \and Luo J.} \REVIEW{Journal of the
  American Chemical Society}{137}{2015}{15660}.

\bibitem{piezo}
\Name{Harada J., Kawamura Y., Takahashi Y., Uemura Y., Hasegawa T., Taniguchi
  H. \and Maruyama K.} \REVIEW{Journal of the American Chemical
  Society}{141}{2019}{9349}.

\bibitem{Das}
\Name{Das S., Mondal A. \and Reddy C.~M.} \REVIEW{Chemical Society
  Reviews}{49}{2020}{8878}.

\bibitem{ammonia}
\Name{Zhang H.} \Book{Experimental investigation of the phase diagram of
  ammonia monohydrate at high pressure and temperature} Ph.D. thesis Sorbonne
  universit{\'e} (2019).

\bibitem{kaneko}
\Name{Kaneko Y. \and Sorai M.} \REVIEW{Phase Transitions}{80}{2007}{517}.

\bibitem{kundu2013nematic}
\Name{Kundu J., Rajesh R., Dhar D. \and Stilck J.~F.} \REVIEW{Physical Review
  E}{87}{2013}{032103}.

\bibitem{jaleel2021rejection}
\Name{Jaleel A. A.~A., Thomas J.~E., Mandal D., Rajesh R. \etal}
  \REVIEW{Physical Review E}{104}{2021}{045310}.

\bibitem{barnes}
\Name{Barnes B.~C., Siderius D.~W. \and Gelb L.~D.}
  \REVIEW{Langmuir}{25}{2009}{6702}.

\bibitem{2018-pre-mnr-phase}
\Name{Mandal D., Nath T. \and Rajesh R.} \REVIEW{Phys. Rev.
  E}{97}{2018}{032131}.

\bibitem{rajesh2}
\Name{Jaleel A. A.~A., Mandal D., Thomas J.~E. \and Rajesh R.} \REVIEW{Physical
  Review E}{106}{2022}{044136}.

\bibitem{rectangles}
\Name{Nath T., Dhar D. \and Rajesh R.} \REVIEW{Europhysics
  Letters}{114}{2016}{10003}.

\bibitem{kundu}
\Name{Kundu J. \and Rajesh R.} \REVIEW{Phys. Rev. E}{89}{2014}{052124}.

\bibitem{geet}
\Name{Mandal D., Rakala G., Damle K., Dhar D. \and Rajesh R.} \REVIEW{arXiv
  preprint arXiv:2109.02611}{}{2021}{}.

\bibitem{rajesh1}
\Name{Vigneshwar N., Mandal D., Damle K., Dhar D. \and Rajesh R.} \REVIEW{Phys.
  Rev. E}{99}{2019}{052129}.

\bibitem{runnels1965hard}
\Name{Runnels L.} \REVIEW{Physical Review Letters}{15}{1965}{581}.

\bibitem{nath2016high}
\Name{Nath T. \and Rajesh R.} \REVIEW{Journal of Statistical Mechanics: Theory
  and Experiment}{2016}{2016}{073203}.

\bibitem{mazel2023hard}
\Name{Mazel A., Stuhl I. \and Suhov Y.} \REVIEW{arXiv preprint
  arXiv:2304.08642}{}{2023}{}.

\bibitem{runnels1}
\Name{Casey L.~M. \and Runnels L.} \REVIEW{The Journal of Chemical
  Physics}{51}{1969}{5070}.

\bibitem{runnels2}
\Name{Freasier B. \and Runnels L.} \REVIEW{The Journal of Chemical
  Physics}{58}{1973}{2963}.

\bibitem{odc1}
\Name{Klamser J.~U., Sadhu T. \and Dhar D.} \REVIEW{Physical Review
  E}{106}{2022}{L052101}.

\bibitem{odc2}
\Name{Saryal S. \and Dhar D.} \REVIEW{Journal of Statistical Mechanics: Theory
  and Experiment}{2022}{2022}{043204}.

\bibitem{odc3}
\Name{Saryal S. \and Dhar D.} \REVIEW{arXiv preprint
  arXiv:2305.14530}{}{2023}{}.

\end{thebibliography}

\end{document}